\documentclass[a4paper,12pt]{article}
\usepackage{theorem, amssymb, amsmath, color, graphics}
\usepackage{mathrsfs}

\DeclareMathOperator{\Ran}{Ran}

\DeclareMathOperator{\Tr}{Tr}

\DeclareMathOperator{\Alg}{Alg}
\DeclareMathOperator{\Sp}{Span}
\DeclareMathOperator{\SL}{SL}
\DeclareMathOperator{\GL}{GL}
\newcommand{\<}{\langle}
\renewcommand{\>}{\rangle}

\def\im{\mathrm{i}}

\def\bbbz{{\mathbb Z}}

\def\bbbc{{\mathbb C}}

\def\Es{\mathbf{E}}

\def\Ps{\mathbf{P}}

\def\iA{{\cal A}}
\def\iB{{\cal B}}
\def\iC{{\cal C}}

\def\iF{{\cal F}}
\def\iS{{\cal S}}

\def\sH{{\mathscr{H}}}

\def\proof{\noindent{\it Proof.} }
\newtheorem{thm}{Theorem}

\newtheorem{prop}{Propostion}
\newtheorem{cor}{Corollary}

\def\qed{{\hfill $\square$}\medskip}

\author{Andr\' as Sz\' ant\' o \thanks{prikolics@gmail.com} \\
\small{Mathematical Institute, Department of Analysis} \\
\small{Budapest University of Technology and Economics}
}
\title{Complementary Decompositions and Unextendible Mutually Unbiased Bases}

\begin{document}

\maketitle

\begin{abstract}
Unextendible sets of Mutually Unbiased Bases (MUBs) are examined from the point of view of complementary subalgebras. We show, that the linear span of less than $d+1$ factors of $M_d \otimes M_d$ does not contain pure states, and therefore some complementary decompositions give rise to undextendable sets of MUBs. We provide some new complementary decompositions, and thus prove strong unextendibility of some set of MUBs.
\end{abstract}

\section*{Introduction}

The concepts of complementarity and mutually unbiased bases (MUBs) arise naturally in quantum information theory. Two orthonormal bases $\{|f_i\>\}_i$ and $\{|g_j\>\}_j$ of a Hilbert space $\sH$ are mutually unbiased if the expression $|\<f_i | g_j\>|^2 = \frac{1}{d}$ is constant regardless of the choice of $i$ and $j$ and depends only on $d=\dim \sH$. It is easy to see, that the maximum number of pairwise unbiased bases is $d+1$. If only von Neumann measurements are used to determine the parameters of an unknown state, then the optimal case is when the measurements are pairwise complementary, that is the eigenbases of the observables are pairwise mutually unbiased \cite{WootersFields:OptStateDetMutUnbiasedBases,Ivonovic:GeomDescQuantStateDet}, and that amounts to a complete set of $d+1$ MUBs. 

There is a long standing conjecture regarding the maximal number of pairwise mutually unbiased bases in dimension $d$, namely that the upper bound $d+1$ is achievable if and only if $d$ is a prime power \cite{Zauner:QDes}. While there are constructions in any prime power dimension \cite{Bandyo:NewProofMUB}, the converse seems to be hard, as even the case $d=6$ is yet to be solved.

A basis of a Hilbert space correspond to the maximal abelian subalgebra of the matrices diagonal in the given basis. Mutually unbiased bases are then equivalent to subalgebras whose traceless parts are orthogonal to each other with respect to the Hilbert-Schmidt scalar product of matrices. The same kind of orthogonality relationship can be studied also for other kind of subalgebras, for example factors are of a special interest \cite{Sajat:QuaOrth4x4, Sajat:CompAlgStr, Weiner:QuaOrthSys}. 

Recently there was some interest in non-complete MUBs that cannot be extended to a complete set \cite{Matolcsi:UEMUB6,Mandayam:UEMUBsPauli,Thas:UEMUBs}. Here we will use complementary decompositions to study the problem. Using the terminology from \cite{Mandayam:UEMUBsPauli}, we construct strongly unextendible MUBs in dimensions $p^2$ where $p$ is any prime such that $p \equiv 3 \pmod 4$ holds. We also show, that the Galois MUBs arise from complementary decompostitions, and are strongly unextendible as well.

In the first section we rehash the basics of the theory and some results on complementarity. In the second section we study the conditional expectation of a pure state with respect to a factor, and develop our tool for proving strong unextendiblity. In the last section, we provide some complementary decompositions and show strong unextendibility of the associated MUBs.

\section{Complementarity}

The choice of a basis $E=\{|e_i\>\}_i$ of the finite dimensional Hilbert space $\sH$ is equivalent to a maximal abelian subalgebra (abbreviated as MASA) $\iC_E$ of the algebra $M_d$ of the matrices acting on the space, consisting of the matrices that are diagonal in the given basis:
$$
\iC_E=\left\{\sum_i \lambda_i |e_i\>\<e_i|: \lambda_i \in \bbbc\right\}
$$
Let us define the normalized trace $\tau:=\frac{1}{d}\Tr$ where $d=\dim(\sH)$. Unbiasedness then can be viewed also as a relation between MASAs. If $F=\{|f_j\>\}_j$ is another basis of $\sH$, then $E$ and $F$ are mutually unbiased
if and only if for any matrices $X \in \iC_E$ and $Y \in \iC_F$ we have
\begin{equation}\label{eq:TrCompl}
\tau(X^*Y)=\tau(X^*)\tau(Y).
\end{equation}

Considering the Hilbert-Schmidt scalar product $\<X, Y\>=\Tr(X^*Y)$ we see, that this relation of MASAs is in fact an 
orthogonality relation of the traceless subspaces $\iC_E \ominus \bbbc I$ and $\iC_F \ominus \bbbc I$, where 
$$
\bbbc I = \{zI | z \in \bbbc\}.
$$
This kind orthogonality can be considered for any pair of 
(not necessarily abelian) subalgebras, and we will call this relation complementarity. We use the term quasi-orthogonality for the same relation between subspaces of any kind, and for any subspace $V$ we define its quasi-orthogonal complement as the orthogonal complement of $V \ominus \bbbc I$. 

We define $\Es_{\iA}$, the trace preserving conditional expectation with respect to the subalgebra $\iA$ as the orthogonal projection to $\iA$. For more details, see \cite{Petz:QuantInfo}.

Parts of the following theorem first appeared in the papers \cite{Popa:OrthPairs} and \cite{Partha:OnEstim}, but there only MASAs were considered. This generalization is from \cite{Sajat:StateTom} and \cite{Sajat:CompAlgStr}.

\begin{thm}\label{thm:equivcomp}
Let $\iA$ and $\iB$ be subalgebras of $M_d$. The following conditions are equivalent:
\begin{enumerate}
\item[(i)] 
The subalgebras $\iA$ and  $\iB$ are complementary, that
is the subspaces $\iA \ominus \bbbc I$ and  $\iB \ominus \bbbc I$ are orthogonal.
\item [(ii)]
$\tau(AB)=\tau(A)\tau(B)$ if $A \in \iA$,  $B \in \iB$.
\item [(iii)]
$\Es_{\iA}(B)=\tau (B) I$ for all $B \in \iB$.
\end{enumerate}
\end{thm}

Since any subalgebra contains the identity matrix $I$, no two subalgebras can be completely orthogonal. In some sense complementarity is the orthogonality of subalgebras to the maximum possible extent.

One of the most important type of subalgebras from the point of view of quantum information theory are factors. A factor is a von Neumann subalgebra that has trivial center i.e. the center is the set of scalar multiples of the identity. In finite dimensional cases this reduces to subalgebras of the form $U (M_d \otimes I) U^*$ where $d$ is a divisor of the dimension of the underlying Hilbert space and $U$ is a unitary matrix. A factor $\iF$ corresponds to a subsystem, as the measurement statistic of any local observable $A=A^*\in \iF$ is completely determined by the conditional expectation $\Es_\iF(\rho)$ of the state $\rho$. For more details, see for example \cite{Petz:QuantInfo}.

A factor $\iF$ of $M_n$ is always complementary to its commutant
$$
\iF'=\{ X\in M_n | \forall A \in \iF: XA=AX\},
$$ 
that is the subalgebra of all matrices commuting with every matrix in $\iF$.
Indeed, $(M_d \otimes \bbbc I)'=\bbbc I \otimes M_\frac{n}{d}$, and for arbitrary traceless matrices $A,B \in M_d$ we have
$$
\Tr \left( (A \otimes I)(I \otimes B)^* \right) = \Tr (A \otimes B^*) =0.
$$
Therefore $M_d \otimes \bbbc I$ and $\bbbc I \otimes M_d$ are complementary, and unitary transformations transform factors together with their commutants. Note, that the commutant of a MASA is itself.

Let $\iF$ be a factor of $M_n$, $\{A_1, \ldots, A_k\}$ an orthonormal basis of $\iF$, and $\{B_1, \ldots, B_l\}$ an orthonormal basis of $\iF'$. Then $\{A_iB_j | i=1, \ldots k, j=1 \ldots l\}$ is an orthonormal basis of $M_n$, and $\iF \otimes \iF' \cong M_n$.

Since a MASA is generated by pairwise orthogonal minimal projections, it is natural to define complementarity of vectors as well, in the following sense: a vector $|v\>$ is complementary to a subalgebra, if the subalgebra generated by the projection $|v\>\<v|$ and the identity is complementary. 

This has physical meaning as well. A pure state $|v\> \in \bbbc^d \otimes \bbbc^n$ of a bipartite quantum system is called entangled, when it is not separable, that is it cannot be written as an elementary tensor $|v_1\> \otimes |v_2\>$. If entanglement is present, then the state of one subsystem cannot be fully described without the other: a measurement on one of the subsystems collapses the other subsystem as well, and the measurement results on the two subsystems are correlated. The correlation is the highest, when the state is maximally entangled, that is the reduced densities are the normalized identity matrices. That is exactly the case of the complementarity of the vector to two factors that are commutant of each other. We will call the state $|v\>$ separable with respect to the factors $\iF$ and $\iF'$, when $|v\>\<v|=PQ$, where $P \in \iF$ and $Q \in \iF'$ are minimal projections. 

The results in the following two propositions are the same as in \cite{Werner:AllTelepDensCodSch}, although there they were phrased quite differently.

Let us fix any pair of bases $\{ |e_i\>\}_i$ of $\bbbc^d$ and $\{ |f_j\>\}_j$ of $\bbbc^n$ and consider the bijection 
$$|v_.\>: M_{d\times n} \to \bbbc^d \otimes \bbbc^n,$$ 
$$|v_A\>=\sum_{i,j} A_{ij} |e_i\> \otimes |f_j\>.$$
Note, that $|v_A\>\<v_A|=\sum_{i,j,k,l} A_{ij} \overline{A_{kl}} \, |e_i\>\<e_k| \otimes |f_j\>\<f_l|$. 

\begin{prop}
 The vector $|v_A\> \in \bbbc^d \otimes \bbbc^n$ is complementary to the subalgebra $M_d \otimes \bbbc I$ if and only if $AA^*=\frac{1}{d}I$.
\end{prop}
\proof
The projection $|v_A\>\<v_A|$ and the matrix $|e_i\>\<e_j| \otimes I$ are complementary iff 
$$\Tr\left(|v_A\>\<v_A|\left(|e_i\>\<e_j| \otimes I\right)\right)=\delta_{ij} \frac{1}{d},$$
and we also have
\begin{align*}
\Tr\left(|v_A\>\<v_A|\left(|e_i\>\<e_j| \otimes I\right)\right)&=\sum_{kl} \overline{A_{ik}}A_{jl} \Tr |e_k\>\<e_l| \\
 &=\sum_{k}\overline{A_{ik}}A_{jk}=(AA^*)_{ji}. \\
\end{align*}
\qed

A similar calculation yields the following

\begin{prop}
 The vector $|v_A\> \in \bbbc^d \otimes \bbbc^n$ is complementary to the subalgebra $\bbbc I \otimes M_n$ if and only if $A^*A=\frac{1}{n}I$
\end{prop}

If $n=d$ then $A$ is constant times a unitary matrix. Also note, that when $n \neq d$, then one of the above conditions on $A^*A$ and $AA^*$ cannot hold. The following simple corollary is a very useful tool in studying the complementary decompositions of $M_4$ \cite{Sajat:CompAlgStr,Sajat:QuaOrth4x4}.

\begin{cor} \label{cor:masasymm}
 Any vector $|v_A\> \in \bbbc^d \otimes \bbbc^d$ -- and so any MASA $\iC \subset M_d \otimes M_d$ -- is complementary to the factor $M_d \otimes \bbbc I$ if and only if it is also complementary to the factor $\bbbc I \otimes M_d$
\end{cor}

A MASA of $M_d \otimes M_d$ complementary to the factors $M_d \otimes I$ and its commutant is the same as a choice of a basis consisting solely of maximally entangled vectors, and the main result of the paper \cite{Werner:AllTelepDensCodSch} states, that this is equivalent to a teleportation scheme, a dense coding scheme, a basis of unitary operators, and a collection of unitary depolarizers.

A set of pairwise complementary subalgebras (factors and MASAs) of $M_d$ is called a complementary decomposition of the algebra $M_d$ when the subalgerbras span $M_d$ linearly. This means, that the direct sum of the traceless subspaces of the subalgebras is $M_d \ominus \bbbc I$. An example would be the MASAs associated with a full set of MUBs when it exists. 

The possible complementary decompositions of $M_2 \otimes M_2$ is well understood, and a complete classification is given in \cite{Sajat:CompAlgStr}. A construction of decomposing $M_{p^{nk}}$ to complementary factors is given by Ohno in \cite{Ohno:QuaOrthSub}. It is also known, that no complementary decomposition of the algebra $M_n \otimes M_n$ to MASAs and factors isomorphic to $M_n$ is possible in which the number of factors is $1$ or $3$ \cite{Weiner:QuaOrthSys}. The existence of any kind of complementary decomposition in not prime power dimensions is still open.

\section{Conditional expectation of a pure state}

We will use the simple fact, that $\Es_{\bbbc I \otimes M_n}(X)= \frac{1}{d} I \otimes \Tr_1 (X)$, where $\Tr_1: M_d \otimes M_n \mapsto M_n$ is the linear operator called partial trace which is defined by the relation $\Tr_1 (A \otimes B) = \Tr(A) B$.

\begin{prop}\label{prp:purepartial}
Let $|h\>\<h| \in M_d \otimes M_n$ be a pure state, and $\iF$ a factor of $M_d \otimes M_n$ isomorphic to $M_n$. Then
$$
\<h|\Es_\iF(|h\>\<h|)|h\> \leq \frac{1}{d},
$$
and there is equality if and only if $|h\>\<h|$ is a separable state with respect to the factors $\iF$ and $\iF'$. 
\end{prop}
\proof
Since all factors are unitary equivalent, we may assume, that $\iF = \bbbc I \otimes M_n$. Let $\{|e_i\>\}_i$ be a basis of $\bbbc^d$, and let us write the state as $|h\>=\sum_i |e_i\> \otimes |h_i\>$. This implies $\sum_i \<h_i|h_i\> = 1$. Then 
\begin{align*}
\Es_\iF(|h\>\<h|) &= I \otimes \frac{1}{d}\Tr_1(|h\>\<h|) \\
&=\frac {1}{d} \sum_i I \otimes |h_i\>\<h_i|
\end{align*}
and
\begin{align*}
\<h|\Es_\iF(|h\>\<h|)|h\> &= \frac {1}{d} \sum_{ijk} \<e_i|e_k\> \otimes \<h_i|h_j\> \<h_j|h_k\> \\
&=\frac {1}{d} \sum_{ij} |\<h_i|h_j\>|^2 \\
& \leq \frac {1}{d} \sum_{ij} \<h_i | h_i\> \<h_j | h_j\> \\
&= \frac {1}{d}
\end{align*}
by the Cauchy-Schwartz inequality. There is equality if and only if the $|h_i\>$ are colinear, that is the state is separable. \qed

As a consequence, we have the following theorem.

\begin{thm}\label{thm:MasaInFactors}
Let $\{\iF_i\}_{i=1}^k$ be a set of pairwise complementary factors of $M_d \otimes M_n$, each isomorphic to $M_n$, such that
there is a pure state in the space $\Sp \{\iF_i | i=1, \ldots k\}$. Then $k\geq d+\frac{d-1}{n-1}$.
\end{thm}
\proof
If $|h\>\<h|$ is in the space $\Sp \{\iF_i | i=1, \ldots k\}$, then 
$$
|h\>\<h| + \frac{(k-1)}{d^2} I = \sum_{i=1}^k \Es_{\iF_i}(|h\>\<h|).
$$
By Proposition \ref{prp:purepartial}, we have
\begin{align*}
1+\frac{(k-1)}{nd} &= \sum_{i=1}^k \<h|\Es_{\iF_i}(|h\>\<h|)|h\> \\
& \leq \frac{k}{d},
\end{align*}
and this is equivalent to  
$$
d+\frac{d-1}{n-1} \le k.
$$
\qed 

As a consequence, the span of less than $d+\frac{d-1}{n-1}$ factors does not contain any MASA, since a MASA is generated by pure states. One could wonder about the reverse situation as well: how many MASAs are needed for their span to contain a factor?

\begin{thm}
Let $\{\iC_i\}_{i=1}^k$ be a set of pairwise complementary MASAs of $M_d \otimes M_n$, and let $\iA \subset \Sp \{\iC_i | i=1, \ldots k\}$ be a factor isomorphic to $M_n$. Then $k \geq n+1$.
\end{thm}
\proof
Let $P_k=|\phi_k \> \< \phi_k |$ for $k=1, \ldots, nd$ be an orthonormal basis of $\iC_i$, so we can write $\Es_{\iC_i}=\sum_k |P_k\>\<P_k|.$ Then by Proposition \ref{prp:purepartial}, 
$$
\< P_k, \Es_{\iA}(P_k)\>= \<\phi_k |\Es_\iF(|\phi_k\>\<\phi_k|)|\phi_k\> \leq \frac{1}{d}
$$
and
$$
\Tr (\Es_{\iA} \Es_{\iC_i}) = \sum_k \< P_k, \Es_{\iA}(P_k)\> \leq n
$$
holds.

Let $\Ps$ be the orthogonal projection to the space $\Sp \{\iC_i | i=1, \ldots k\}$, so 
$$
\Ps=\Es_{\bbbc I} + \sum_i (\Es_{\iC_i} - \Es_{\bbbc I}),
$$
and since $\iA \subset \Sp \{\iC_i | i=1, \ldots k\}$,
$$
\Es_{\iA}=\Es_{\iA} \Ps.
$$
Then we have the following bound: 
\begin{align*}
n^2 &= \Tr \Es_{\iA} \\
&=\Tr (\Es_{\iA} \Ps) \\
&= \Tr (\Es_{\iA} \Es_{\bbbc I}) + \Tr \left( \Es_{\iA} \sum_i (\Es_{\iC_i} - \Es_{\bbbc I})\right) \\
&\leq 1 + \sum_i (n-1) \\
&=1+k(n-1), \\
\end{align*}
and that reduces to
$$
k \geq n+1.
$$
\qed

We have an interesting application of Theroem \ref{thm:MasaInFactors} to unextendible MUBs. A set of MUBs is called unextendible, if there is no basis mutually unbiased with all of the bases in the set. We use the term strongly unextendible, if there is not even a single vector unbiased with respect to the bases \cite{Mandayam:UEMUBsPauli}. 

In this paper we only consider the case $n=d$, and then the inequality in the theorem reduces to $k \geq d+1$. We intend to examine the case when $nd$ is a prime power in a forthcoming paper. If we have a complementary decomposition with factors less than $d+1$, then the quasi-orthogonal complement of the MASAs in the decomposition does not contain any pure state. Therefore the corresponding MUBs are strongly unextendible, so we have the following corollary.

\begin{cor} \label{cor:unextendible}
Let $\{\iF_i\}_{i=1}^k \cup \{\iC_i\}_{i=1}^{d^2+1-k}$ with $k < d+1$ be a complementary decomposition of $M_d \otimes M_d$, where $\iF_i$ are factors isomorphic to $M_d$, and the subalgebras $\iC_i$ are MASAs. Then the set of MUBs corresponding to the MASAs is strongly unextendible.
\end{cor}

It would be interesting to see if the converse is true, that is if there is any set of (strongly or not strongly) unextendible MASAs such that the quasi-orthogonal complement is not decomposable to pairwise complementary factors.

\section{Complementary decompositions}

Let $p \geq 2$ be a prime. We will construct $p^2-1$ subalgebras that together with the factors $M_p \otimes \bbbc I$ and $\bbbc I \otimes M_p$ they form a complementary decomposition of $M_p \otimes M_p$.

The following construction of a basis consisting only unitary matrices was originally considered by Sylvester \cite{Sylvester:Nonions}, and was also studied by Schwinger \cite{Schwinger:UnOpBases}. The matrices are sometimes called as shift and clock matrices, and also generalized Pauli matrices.

 Consider an orthonormal basis $\{|e_i\>\}_i$ of the Hilbert space $\sH$ with dimension $p$, and the root of unity $\omega = e^{\frac{2\im\pi}{p}}$. Then we define the matrices
$$
Z=\sum_{i=1}^p \omega^i |e_i\>\<e_i|
$$
and
$$
X=\sum_{i=1}^p |e_{i+1}\>\<e_i|,
$$
where the addition in the index of the basis vectors is understood modulo $d$.

It is easy to see, that the matrices $\{X^iZ^j\}_{i,j=1}^p$ are unitary and form an orthonormal basis of $M_p$. We also have the commutation relation
\begin{equation} \label{eq:SchwingerCommutation}
XZ=\frac{1}{\omega} ZX, 
\end{equation}
which makes computation in this basis convenient. These matrices are closely related to the Pauli matrices: for $p=2$ we have $Z=-\sigma_3$ and $X=\sigma_1$.

Let V be the 4-dimensional vectorspace over $\bbbz_p$, and let $u,v \in V$. Define the map $\pi$ as
$$
\pi(u,v) = \Alg \{ X^{u_1}Z^{u_2} \otimes X^{u_3}Z^{u_4}, X^{v_1}Z^{v_2} \otimes X^{v_3}Z^{v_4} \},
$$
or equivalently for a 2-dimensional subspace $U < V$ as
$$
\pi(U) = \Alg \{X^{u_1}Z^{u_2} \otimes X^{u_3}Z^{u_4} | u \in U \}.
$$
We say that two such subspaces are complementary, if their images under $\pi$ are complementary subalgebras, or equivalently if their intersection is trivial. Consider the antisymmetric bilinear form
$$
c(u,v)=u_1 v_2 - u_2 v_1 + u_3 v_4 - u_4 v_3.
$$
Note, that $c(u, ku)=0$ for all $u\in V, k \in \bbbz_p$, and for any 2-dimensional subspace $A < V$ 
\begin{itemize}
\item either $c(u,v)=0$ for any $u, v \in A$ (then $\pi(u,v)$ is a MASA), 
\item or $c(u,v)=0$ for $u,v \in V$ if and only if $u=kv$ for some $k\in \bbbz_p$ (and in this case $\pi(u,v)$ is a factor).
\end{itemize}

Denote with $\iS$ the 2-dimensional subspaces $S < V$ such that $\pi(S)$ is complementary to $M_p \otimes \bbbc I$ and to its commutant. Note, that $S \in \iS$ if and only if $S$ has trivial intersection with the subspaces
$$F_0=\Sp\{(1,0,0,0),(0,1,0,0)\}$$
and
$$F_1=\Sp\{(0,0,1,0),(0,0,0,1)\}.$$

Consider the map 
$$
\phi\left(\left(\begin{array}{cc} x_1 & y_1 \\ x_2 & y_2 \end{array}\right)\right) = \Sp \{(0,1,x_1,x_2), (1,0,y_1,y_2)\}.
$$

\begin{prop}
The map $\phi$ is a bijection $\GL_2(p) \to \iS$.
\end{prop}
\proof
 Let $M\in GL_2(p)$.

 First, we show that $\Ran \phi \subseteq \iS$. If the columns of the matrix $M$ are u and v, then 
$$
\phi(M)=\{ (k, l, k u_1+l v_1, k u_2 + l v_2 | k,l \in \bbbz_p \}.
$$ 
Since $u$ and $v$ are linearly independent, $k=l=0$ and $k u_1+l v_1 = k u_2 + l v_2 = 0$ can only happen at the same time, so  $\phi(M) \cap F_0=\phi(M) \cap F_1 = \{0\}$.

It is clear, that $\phi$ is injective. We show that $\phi$ is surjective. Let $S \in \iS$. There are no two vectors in $S$ that are independent but whose projections to the first two coordinates are dependent, as then a nontrivial linear combination would be in $F_1$. We can then reduce any basis of $S$ to a row echelon form $(1,0,x,y)$ and $(0,1,w,z)$. The vectors $(x,y)$ and $(w,z)$ must be independent, otherwise $S \cap F_0$ would not be trivial. We arrive at the representing matrix 
$$
M_S=\left(\begin{array}{cc} x & y \\ w & z \end{array}\right)
$$
with $\phi(M_S)=S$ \qed

For brevity, when $M\in \GL_2(p)$ the expression $\pi(M)$ is used instead of $\pi(\phi(M))$. Note, that a subalgebra is a MASA exactly when the determinant of the representing matrix $M$ is 1, that is $M\in \SL_2(p)$, otherwise it is a factor. The determinant also appears in the computation of the commutant.

\begin{prop}
If $M\in \GL_2(p)$, then
$$
\pi(M)'=\pi\left(\frac{1}{\det M} M \right).
$$
\end{prop}
\proof
Let 
$$
M=\left(\begin{array}{cc} x_1 & y_1 \\ x_2 & y_2 \end{array}\right).
$$ 
Then the subspace $\phi(M)$ is generated by the vectors
$$u=(0,1,x_1,y_1),$$ 
$$v=(1,0,x_2,y_2)$$ 
and we have $c(u,v)=1-\det(M)$. 

Multiplying one row with $\frac{1}{\det M}$ multiplies the determinant as well, so for the vectors
$$u'=(0,1 ,\frac{1}{\det M} x_1, \frac{1}{\det M} y_1),$$ 
$$v'= (1,0, \frac{1}{\det M} x_2, \frac{1}{\det M} y_2)$$ 
we have
$c(u,u')=c(u,v')=c(v,u')=c(v,v') = 0$. Since $\phi\left(\frac{1}{\det M} M \right)$ is generated by $u'$ and $v'$, the assertion must hold. \qed

\subsection{Galois MUBs}

Here we construct decompositions with the two product factors and $p^2-1$ MASAs. The proof of the next proposition relies on the fact, that for any $2 \times 2$ matrix $M$ the equation
\begin{equation} \label{eq:2x2mat}
M^2-(\Tr M) M + (\det M) I = 0
\end{equation}
holds. 

\begin{prop} \label{prp:SL2MASA}
Let $A,B \in \SL_2(p)$. Then the following are equivalent.
\begin{enumerate}
 \item[(i)] The intersection of $\phi(A)$ and $\phi(B)$ is not trivial.
 \item[(ii)] There are $u,v \in \bbbz_p$ not both zero such that 
   $$
	 (A - B) \left( \begin{array}{c} u \\ v \end{array} \right) = 0.
	 $$
 \item[(iii)] $\det (A - B) = 0$
 \item[(iv)] $\det (A^{-1}B - I)=0$
 \item[(v)] $A^{-1}B$ is either the identity matrix or it has order $p$.
\end{enumerate}
\end{prop}
\proof The equivalences of (i)-(iv) are trivial. 

Let $M=A^{-1}B$.

Assume that (iv) holds, and $M\neq I$. Then $\det M=1$ and $\det (M-I)=0$, so substituting $M$ and $M-I$ into equation (\ref{eq:2x2mat}) and solving for $\Tr M$ yields, that $\Tr M=2$ and $M^2-2M+I=0$. If $p=2$ then $M^2=I$, otherwise $(x-1)^2$, the minimal polynomial of $M$ divides $(x-1)^p \equiv x^p-1 \pmod p$, so $M^p-I=0$. Since $p$ is prime, the order of $M$ must be $p$.

Now assume (v). Then $\Tr M^p=2$. Let $t_k=\Tr M^k$, then by multiplying (\ref{eq:2x2mat}) by $M^k$ and taking the trace we see that the recurrence relation $t_{k+2}-t_1t_{k+1}+t_k=0$ holds, with the further condition $t_p=2$. The only solution is $t_1=\Tr M=2$. Substituting $M-I$ into equation (\ref{eq:2x2mat}) shows that $\det(M-I)=0$. \qed

A natural idea is to try to find a subgroup $H\leq \SL_2(p)$ of order $(p+1)(p-1)$, since then there is no element of order $p$ in $H$, as the order of an element divides the order of the group. In the paper \cite{Thas:UEMUBs} these constructions are called Galois MUBs. A similiar theorem as Proposition \ref{prp:SL2MASA} and the same MUBs also appear in \cite{Howard:BipEntStabMUB}. 

For $p=2$, we have the group generated by identity and the matrices 
$$\left(\begin{array}{cc} 1 & 1 \\ 0 &  1 \end{array}\right)$$ 
and 
$$\left(\begin{array}{cc} 1 & 0 \\ 1 & 1 \end{array}\right).$$ 
For $p=3$ the $2$-Sylow subgroup does the job. It is generated by the matrices 
$$\left(\begin{array}{cc} 0 & 1 \\ 2 &  0 \end{array}\right)$$ 
and 
$$\left(\begin{array}{cc} 2 & 2 \\ 2 &  1 \end{array}\right),$$ 
and it is isomorphic to the quaternion group. For $p=5$ we have the normalizer of a 2-Sylow subgroup, it is isomorphic to $\SL_2(3)$. There is also such a subgroup for $p=7,11$, for the generator matrices see \cite{Howard:BipEntStabMUB}. Unfortunately there is no such subroup for greater primes (See for example \cite{Suzuki:GroupTheoryI}). 

The MUBs give rise to a complementary decomposition with two factors, so Corollary \ref{cor:unextendible} applies, and we have proven the following.

\begin{thm}
The Galois MUBs in dimensions $p^2$ with $p=2,3,5,7,11$ are strongly unextendible.
\end{thm}

To the best of our knowledge, this result is new.

\subsection{Decomposition with $p\pm 1$ factors}

The following we use some ideas from the decomposition in \cite{Ohno:QuaOrthSub}. Let $p>2$, fix $D \in \bbbz_p$ a quadratic non-residue, and let us define for $i\in \bbbz_p$ and $0 \neq j \in \bbbz_p$ the matrices
$$
A_{i,j}=\left(\begin{array}{cc} i & -j \\ j^{-1}(1-Di^2) & Di \end{array}\right)
$$

Clearly all these matrices are in $\SL_2(p)$, so let $V_{i,j}=\phi(A_{i,j})$.

\begin{prop} 
The matrices $A_{i,j}$ are all different, and the subspaces $V_{i,j}$ are pairwise complementary.
\end{prop}

\proof We have to show, that $\det(A_{i,j}-A_{x,y}) =0 $ if and only if $i=x$ and $j=y$. Calculating the determinant after substituting $u=yj^{-1}$ yields
$$
Du^{-1}(ui-x)^2 - u^{-1}(u-1)^2=0,
$$
and by the condition on $D$ we conclude, that $u=1$ and $i=x$.
\qed

Since the cardinality of the set of $A_{i,j}$ matrices is only $p(p-1)$, this does not constitute yet a full decomposition. Consider the matrices
$$
B_i=\left(\begin{array}{cc} i & 0 \\ 0 & -iD \end{array}\right),
$$
where $i \neq 0$, and let $V_{i,0}=\phi(B_i)$. Direct calculation shows, that $\det(B_i - B_j) \neq 0$ if $i\neq j$, and
$$
\det(A_{i,j}-B_x)=1-Dx^2\neq 0.
$$

The type of the algebras $\pi(B_i)$ depends on $p$ modulo 4.
We have $\det B_i = -Di^2 = 1$ for some $i$ if and only if $-1$ is a quadratic non-residue, that is when $p=4k+3$. In this case $-D^{-1}$ has exactly two square-roots. Let us denote them $q$ and $p-q$. Then exactly two of the subalgebras $\pi(B_i)$ are abelian: $\pi(B_q)$ and $\pi(B_{p-q})$. 

If $p=4k+1$, then all subalgebras $\pi(B_i)$ are factors.

Note, that in both cases, for all factors among the $\pi(B_i)$, their commutant is also there.

\begin{thm}
The set of subalgebras defined by the matrices $A_{i,j}$ and $B_j$ for $i\in \bbbz_p$ and $0 \neq j \in \bbbz_p$ together with $M_p \otimes I$ and $I \otimes M_p$ form a complementary decomposition of $M_p \otimes M_p$. The number of factors in the decomposition is $p-1$ if $p \equiv 3 \pmod 4$ and $p+1$ otherwise.
\end{thm}

Then using Corollary \ref{cor:unextendible} we arrive at the following.

\begin{cor}
Let $p$ be a prime such that $p \equiv 3 \pmod 4$, and let $q^2 \equiv -D^{-1} \pmod p$. Then the set of MUBs associated with the MASAs defined by the matrices $B_q$, $B_{p-q}$ and $A_{i,j}$ for $i\in \bbbz_p$ and $0 \neq j \in \bbbz_p$ is strongly unextendable.
\end{cor}

The MUBs are always extendible for $p \equiv 1 \pmod 4$, as it is easy to see that the union of the subspaces 
$$\Sp \{(1,0,0,0), (0,1,0,0)\},$$ 
$$\Sp \{(0,0,1,0),(0,0,0,1)\}$$ 
and 
$$\phi(B_i)$$ 
for nonzero i is the same as the union of the subspaces 
$$\Sp\{(1,0,0,0),(0,0,1,0)\},$$ 
$$\Sp\{(0,1,0,0),(0,0,0,1)\}$$ 
and 
$$\Sp\{(1,i,0,0),(0,0,1,-iD)\}.$$ 
The latter $p+1$ subspaces all correspond to pairwise complementary MASAs. This shows, that the MASAs corresponding to the matrices $A_{i,j}$ are always extendible. Also, in the case $p \equiv 3 \pmod 4$ the unextendibility depends on the MASAs corresponding to $B_q$, $B_{p-q}$.

\section{Conclusion}

We showed, that the linear span of less than $d+1$ factors of $M_d \otimes M_d$ does not contain pure states. This implies, that complementary decompositions with less than $d+1$ factors give rise to strongly undextendable sets of MUBs. One of such sets are the so called Galois MUBs. We also construct a complementary decomposition of $M_p \otimes M_p$ with the number of factors depending on $p \, \mathrm{mod}\, 4$, and if $p \equiv 3 \pmod 4$, the MASAs correspond to a set of strongly unextendlible MUBS.

\section*{Acknowledgements}
I would like to thank Markus Grassl for his insightful questions and remarks.

\bibliographystyle{prikolics}
\bibliography{prikolics}

\end{document}